\documentclass[twocolumn,showpacs,prd]{revtex4}
\usepackage{mathrsfs}
\usepackage{longtable,lscape}
\usepackage{txfonts}
\usepackage{amssymb}
\usepackage{indentfirst}
\usepackage{graphicx,,booktabs}
\usepackage{color}
\usepackage{amssymb}

\begin{document}
%
%
\title{Interpretation of $Z_b(10610)$ and $Z_b(10650)$ in the ISPE mechanism and the Charmonium Counterpart}
%
%
\author{Dian-Yong Chen$^{1,3}$}\email{chendy@impcas.ac.cn}
\author{Xiang Liu$^{1,2}$\footnote{corresponding author}}\email{xiangliu@lzu.edu.cn}
\author{Takayuki Matsuki$^4$}\email{matsuki@tokyo-kasei.ac.jp}
\affiliation{
$^1$Research Center for Hadron and CSR Physics,
Lanzhou University and Institute of Modern Physics of CAS, Lanzhou 730000, China\\
$^2$School of Physical Science and Technology, Lanzhou University, Lanzhou 730000,  China\\
$^3$Nuclear Theory Group, Institute of Modern Physics of CAS, Lanzhou 730000, China\\
$^4$Tokyo Kasei University, 1-18-1 Kaga, Itabashi, Tokyo 173-8602, Japan}
\date{\today}

\date{\today}

\begin{abstract}
The initial single pion emission (ISPE) mechanism is applied to the
processes $\Upsilon(5S)\to \pi B^{(*)}\bar{B}^{(*)}$ whose details
have been recently reported at ICHEP2012 and we obtain reasonable
agreement with Bell's measurements, i.e., we succeed in reproducing
the enhancement structures of $Z_b(10610)$ and $Z_b(10650)$.
Inspired by this success, we predict the corresponding
enhancement structures in higher charmonia open charm pion decay
near the thresholds of $D^\ast \bar{D}$ and $D^\ast \bar{D}^\ast$.
\end{abstract}

\pacs{13.25.Gv, 14.40.Pq, 13.75.Lb}
\maketitle

\section{Introduction}\label{sec1}

Two charged bottomonium-like structures $Z_b(10610)$ and
$Z_b(10650)$ were reported by the Belle Collaboration in the
hidden-bottom decays of $\Upsilon(5S)$ \cite{Belle:2011aa}. As
indicated by the analysis of the corresponding $\pi^\pm
\Upsilon(nS)$ $(n=1,2,3)$ and $\pi^\pm h_b(mP)$ $(m=1,2)$ invariant
mass spectra, $Z_b(10610)$ and $Z_b(10650)$ are two narrow
structures with masses and widths $M_{Z_b(10610)}=(10607.2\pm2.0)$
MeV, $\Gamma_{Z_b(10610)}=(18.4\pm2.4)$ MeV,
$M_{Z_b(10650)}=(10652.2\pm1.5)$ MeV, and
$\Gamma_{Z_b(10650)}=(11.5\pm2.2)$ MeV \cite{Belle:2011aa}. In
addition, the spin-parity quantum numbers are $J^P=1^+$ both for
$Z_b(10610)$ and $Z_b(10650)$ due to the analysis of charged pion
angular distribution \cite{Collaboration:2011gja}.

Observation of these two structures has inspired theorists with the
extensive interests. Various theoretical explanations were proposed
after the Belle's observation. In the following, we will briefly
review the research status of $Z_b(10610)$ and $Z_b(10650)$.

Considering that $Z_b(10610)$ and $Z_b(10650)$ are charged and close
to $B\bar{B}^*$ and $B^*\bar{B}^*$ thresholds, respectively, many
theoretical efforts have been made to answer the question whether
these newly observed structures are the real exotic states or not.
Before the discovery of $Z_b(10610)$ and $Z_b(10650)$, the authors
in Refs. \cite{Liu:2008fh,Liu:2008tn} predicted the existence of
loosely bound S-wave $B\bar{B}^*$ molecular states. The heavy quark
spin structure by Bondar {\it et al.} \cite{Bondar:2011ev},
study using the chiral constituent quark model in Ref.
\cite{Yang:2011rp}, the effective Lagrangian approach via the
one-boson exchange in Ref. \cite{Sun:2011uh}, and
{study on the line shape in the vicinity of $B^{(*)}\bar B^{(*)}$ thresholds
as well as two-body decay rates using the effective field theory in Ref. \cite{Mehen:2011yh},}
all showed that
$Z_b(10610)$ and $Z_b(10650)$ can be the $B\bar{B}^*$ and
$B^*\bar{B}^*$ molecular states, respectively. The authors in Ref.
\cite{Yang:2011rp} further showed that their quantum numbers are
$I(J^{PC})=1(1^{+-})$.
%
%
The QCD sum rule (QSR) analysis by Zhang {\it et al.}
\cite{Zhang:2011jja} suggested that $Z_b(10610)$ could be a
$B\bar{B}^*$ molecular state.
%
%
Using the Bethe-Salpeter equation, the problem whether $Z_b(10610)$
is a $B\bar{B}^*$ molecular state was studied in Ref.
\cite{Ke:2012gm}. They claimed that $B\bar{B}^*$ molecular state
with isospin $I=1$ cannot be formed when the contribution of
$\sigma$ exchange is small \cite{Ke:2012gm}.

Apart from these studies of mass spectrum just mentioned above,
there are some theoretical papers dedicated to the production and
decay behavior of $Z_b(10610)$ and $Z_b(10650)$. Under the
frameworks of $B\bar{B}^*$ and $B^*\bar{B}^*$ molecular states, the
radiative decay of $\Upsilon(5S)$ into molecular bottomonium was
calculated \cite{Voloshin:2011qa}, and the processes of
$Z_{b}(10610)$ and $Z_b(10650)$ decaying into bottomonium and pion
were also investigated very recently \cite{Li:2012uc}. In Ref.
\cite{Cleven:2011gp}, the properties of $Z_b(10610)$ and
$Z_b(10650)$ were studied assuming that $Z_b(10610)$ and
$Z_b(10650)$ are the $B\bar{B}^*$ and $B^*\bar{B}^*$ molecular
states. Dong {\it et al.} \cite{Dong:2012hc} performed the
calculation of molecular hadrons, $Z_{b}(10610)$ and $Z_b(10650)$,
decaying into $\Upsilon(nS)$ and $\pi^+$ by the effective Lagrangian
approach.

In addition, tetraquark explanation for $Z_b(10610)$ and
$Z_b(10650)$ was proposed. In Ref. \cite{Guo:2011gu}, the masses of
tetraquark states $bu\bar{b}\bar{d}$ and $bd\bar{b}\bar{u}$ with
$J^P=1^+$ were obtained by the chromomagnetic interaction
Hamiltonian, which are compatible with the corresponding masses of
$Z_b(10610)$ and $Z_b(10650)$. Using the QSR approach, the authors
in Ref. \cite{Cui:2011fj} calculated the mass of the tetraquark
states with the configuration $[bd][\bar{b}\bar{u}]$ and found that
$Z_b(10610)$ and $Z_b(10650)$ can be described by tetraquark. Ali
{\it et al.} also gave tetraquark interpretation for $Z_b(10610)$
and $Z_b(10650)$ and studied the decay of tetraquark state
$Y_b(10890)$ into $Z_b(10610)^\pm \pi^\mp$ or $Z_b(10650)^\pm
\pi^\mp$, and the decays of $Z_b(10610)/Z_b(10650)$ into $\pi^\pm
\Upsilon(nS)$ and $\pi^\pm h_b(mP)$ \cite{Ali:2011ug}.

Besides proposing exotic states to understand these structures,
theorists also tried to explain why $Z_b(10610)$ and $Z_b(10650)$
were observed in the hidden-bottom decays of $\Upsilon(5S)$. Bugg
suggested that two observed structures of $Z_b(10610)$ and
$Z_b(10650)$ are due to cusp effects \cite{Bugg:2011jr}. The authors
in Ref. \cite{Chen:2011zv} indicated newly observed $Z_b(10610)$ and
$Z_b(10650)$ play an important role to describe Belle's previous
observation of the anomalous $\Upsilon(2S)\pi^+\pi^-$ production
near the peak of $\Upsilon(5S)$ at $\sqrt s=10.87$ GeV
\cite{Abe:2007tk}, where the resulting distributions,
$d\Gamma(\Upsilon(5S)\to \Upsilon(2S)\pi^+\pi^-)/dm_{\pi^+\pi^-}$
and $d\Gamma(\Upsilon(5S)\to \Upsilon(2S)\pi^+\pi^-)/d\cos\theta$,
agree with Belle's measurements after inclusion of these $Z_b$
states \cite{Chen:2011zv}. Later, the initial single pion emission
(ISPE) mechanism was proposed in the $\Upsilon(5S)$ hidden-bottom
dipion decays, where the line shapes of $d\Gamma(\Upsilon(5S\to
\Upsilon(nS)\pi^+\pi^-))/dm_{\Upsilon(nS)\pi^+}$ ($n=1,2,3$) and
$d\Gamma(\Upsilon(5S\to h_b(mP)\pi^+\pi^-))/dm_{h_b(mP)\pi^+}$
($m=1,2$) are given \cite{Chen:2011pv}. The sharp structures
obtained around 10610 MeV and 10650 MeV in the theoretical line
shapes of distributions, $d\Gamma(\Upsilon(5S\to
\Upsilon(nS)\pi^+\pi^-))/dm_{\Upsilon(nS)\pi^+}$ and
$d\Gamma(\Upsilon(5S\to h_b(mP)\pi^+\pi^-))/dm_{h_b(mP)\pi^+}$,
could naturally correspond to the $Z_b(10610)$ and $Z_b(10650)$
structures newly observed by Belle \cite{Chen:2011pv}.

Although there have been many theoretical efforts to clarify
$Z_b(10610)$ and $Z_b(10650)$, further study on these two $Z_b$
states is still an interesting research topic. For instance, it is
crucial how to distinguish different explanations for $Z_b(10610)$
and $Z_b(10650)$. Very recently, the Belle Collaboration has
reported new results on $Z_b(10610)$ and $Z_b(10650)$ at the
ICHEP2012 conference that these $Z_b$ structures also exist in the
$B\bar{B}^*$ and $B^*\bar{B}^*$ invariant mass spectra of
$\Upsilon(5S)\to \pi B\bar{B}^*,\,\pi B^*\bar{B}^*$ decays
\cite{newbelle}. This new experimental phenomenon of $Z_b(10610)$
and $Z_b(10650)$ can provide an important platform to test
explanations for $Z_b(10610)$ and $Z_b(10650)$ proposed so far and
this process also reminds us the ISPE mechanism.

In this work, we will explain why two charged structures
$Z_b(10610)$ and $Z_b(10650)$ can appear in the $B\bar{B}^*$ and
$B^*\bar{B}^*$ invariant mass spectra of the $\Upsilon(5S)\to \pi
B\bar{B}^*,\,\pi B^*\bar{B}^*$ decays. We find that the ISPE
mechanism proposed in Ref. \cite{Chen:2011pv} can be well applied to
the $\Upsilon(5S)\to \pi B\bar{B}^*,\,\pi B^*\bar{B}^*$ processes,
which can further test this mechanism.
Other than explaining the Belle's new observation, we will extend
our study to the open-charm decays of higher charmonia with the
emission of a single pion because of the similarity between
bottomonium and charmonium \cite{Chen:2011xk}. As a result of our
study, we will give the corresponding prediction of two charged
charmonium-like structures close to the $D^* \bar{D}$ and
$D^*\bar{D}^*$ thresholds, which can be found in the
invariant mass spectra $m_{D^*\bar D}$ and $m_{D^*\bar D^*}$
of the open-charm decays
of higher charmonia with the emission of a single pion.

This work is organized as follows. After introduction, we introduce
the ISPE mechanism and its application to $\Upsilon(5S)\to \pi
B\bar{B}^*,\,\pi B^*\bar{B}^*$ decays in the next section. The
relevant numerical results will be presented here. In Sec.
\ref{sec3}, we extend the ISPE mechanism to study the open-charm
decays of higher charmonia with the emission of a single pion and
give the corresponding prediction. The paper ends with summary in
Sec. \ref{sec4}.

\section{The ISPE mechanism and the $\Upsilon(5S)\to \pi B^{(*)}\bar{B}^{(*)}$ decays}\label{sec2}

The ISPE mechanism has been first proposed to study the
$\Upsilon(5S)\to\Upsilon(nS)\pi^+\pi^-$ $(n=1,2,3)$ and
$\Upsilon(5S)\to h_b(mP)\pi^+\pi^-$ $(m=1,2)$ decays
\cite{Chen:2011pv}, and it explains why two $Z_b$ structures can be
observed in these processes. Via the ISPE mechanism, the
hidden-bottom dipion decays of $\Upsilon(5S)$ can occur through two
steps. First, $\Upsilon(5S)$ decays into the $B^{(*)}\bar{B}^{(*)}$
plus one pion, where most of the kinematical energy is carried out
by the emitted pion and is continuously distributed.
Secondly, the $B^{(*)}$ and $\bar{B}^{(*)}$ mesons with low momentum
can easily interact with each other to convert into the final state
$\Upsilon(nS)\pi$ or $h_b(mP)\pi$ via the $B^{(*)}$ meson exchange
\cite{Chen:2011pv}.

\begin{figure}[htb]
\centering%
\begin{tabular}{ccc}
\scalebox{0.69}{\includegraphics{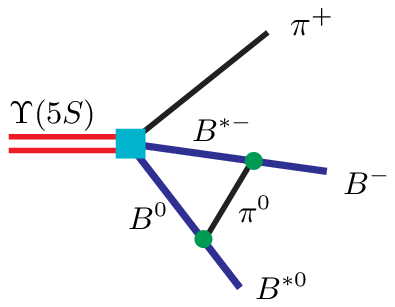}}&%
\scalebox{0.69}{\includegraphics{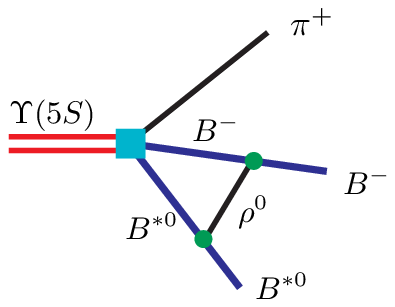}}&%
\scalebox{0.69}{\includegraphics{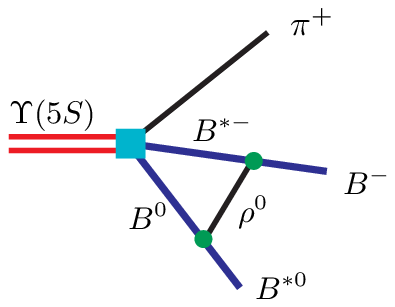}}\\%
(a) & (b) & (c)\\
\scalebox{0.69}{\includegraphics{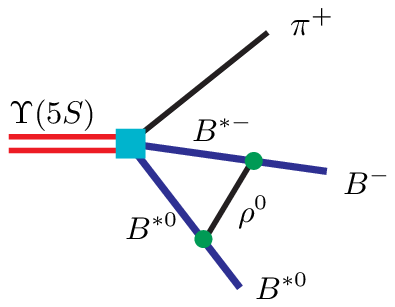}}&%
\scalebox{0.69}{\includegraphics{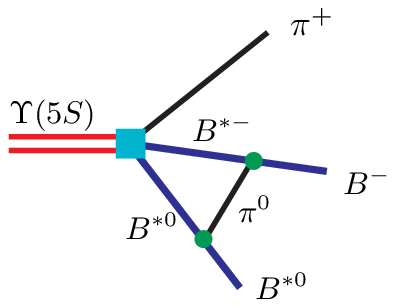}}&\\%
(d) & (e) &
\end{tabular}
\caption{(Color online.) The typical diagrams for the $\Upsilon(5S)
\to B^{*0} B^- \pi^+$ decays via the ISPE mechanism. \label{Fig:ADia}}
\end{figure}

\begin{figure}[htb]
\centering%
\begin{tabular}{ccc}
\scalebox{0.8}{\includegraphics{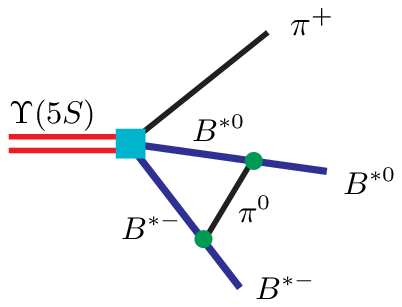}}&%
\scalebox{0.8}{\includegraphics{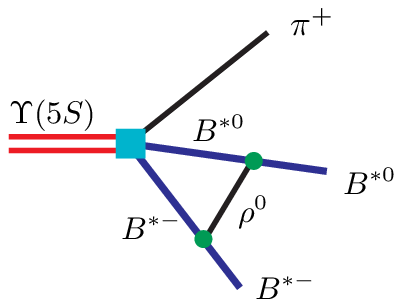}}\\%
(1) & (2)
\end{tabular}%
\caption{(Color online.) The schematic diagrams for the $\Upsilon(5S)
\to B^{\ast0} B^{\ast -} \pi^+$ decays by the ISPE mechanism. \label{Fig:BDia}}
\end{figure}

In this paper, we would like to apply the ISPE mechanism to the
open-bottom decays of $\Upsilon(5S)$ with the emission of a single
pion. In Figs. \ref{Fig:ADia} and \ref{Fig:BDia}, we present the
typical diagrams describing $\Upsilon(5S)\to \pi
B^{(*)}\bar{B}^{(*)}$ via the ISPE mechanism, where the intermediate
$B^{(*)}$ and $\bar{B}^{(*)}$ meson can convert into $B\bar{B}^*$ or
$B^*\bar{B}^*$ final state by exchanging light mesons such as $\pi$
and $\rho$.

In the following, we will write out the decay amplitudes corresponding to the diagrams listed in Figs. \ref{Fig:ADia} and \ref{Fig:BDia}.
The effective Lagrangians relevant to our study are given by
\begin{eqnarray}
\mathcal{L}_{\Upsilon(5S) \mathcal{B}^{(*)} \mathcal{B}^{(*)} \pi}
&=& -ig_{\Upsilon^\prime \mathcal{B}\mathcal{B} \pi} \varepsilon^{\mu
\nu \alpha \beta} \Upsilon_{\mu} \partial_{\nu} \mathcal{B}
\partial_{\alpha} \pi \partial_{\beta} \bar{\mathcal{B}} \nonumber\\
&&+ g_{\Upsilon \mathcal{B}^\ast \mathcal{B} \pi}
{\Upsilon}^{\mu} (\mathcal{B} \pi \bar{\mathcal{B}}^\ast_{\mu} +
\mathcal{B}^\ast_{\mu} \pi \bar{\mathcal{B}})\nonumber\\&&  -ig_{\Upsilon
\mathcal{B}^\ast \mathcal{B}^\ast \pi} \varepsilon^{\mu \nu \alpha
\beta} \Upsilon_{\mu} \mathcal{B}^\ast_{\nu} \partial_{\alpha} \pi
\bar{\mathcal{B}}^\ast_\beta \nonumber\\
&&-ih_{\Upsilon \mathcal{B}^\ast \mathcal{B}^\ast \pi}
\varepsilon^{\mu \nu \alpha \beta}
\partial_{\mu} \Upsilon_{\nu}
\mathcal{B}^\ast_{\alpha} \pi \bar{\mathcal{B}}^\ast_{\beta},\\
\mathcal{L}_{\mathcal{B}^{(\ast)} \mathcal{B}^{(\ast)}
\mathcal{V}}&=& -i g_{\mathcal{BBV}}\bar{\mathcal{B}}
{\stackrel{\leftrightarrow}{\partial}}_\mu \mathcal{B}
(\mathcal{V}^\mu) \nonumber\\&&-2f_{\mathcal{B B^\ast V}}
\epsilon_{\mu\nu\alpha\beta}(\partial^\mu \mathcal{V}^\nu)
 (\bar{\mathcal{B}}
{\stackrel{\leftrightarrow}{\partial}^\alpha} \mathcal{B}^{*\beta} -
\bar{\mathcal{B}}^{\ast \beta}
{\stackrel{\leftrightarrow}{\partial}^\alpha} \mathcal{B}) \nonumber\\&&+ i
g_{\mathcal{B^\ast B^\ast V}} \bar{\mathcal{B}}^{\ast \nu}
{\stackrel{\leftrightarrow}{\partial}_\mu} \mathcal{B}^\ast_{\nu}
(\mathcal{V}^\mu) \nonumber\\ && + 4i f_{B^\ast B^\ast V}
\bar{\mathcal{B}}^{\ast \mu} (\partial_\mu \mathcal{V}_\nu -
\partial_\nu \mathcal{V}_\mu) \mathcal{B}^{\ast \nu},
\end{eqnarray}
\begin{eqnarray}
\mathcal{L}_{\mathcal{B}^\ast \mathcal{B}^{(\ast)} \mathcal{P}} &=&
-i g_{\mathcal{B^\ast B V}} (\bar{\mathcal{B}}_i
\partial_\mu \mathcal{P}_{ij} \mathcal{B}^{*\mu}_j -
\bar{\mathcal{B}}^{*\mu}_i \partial_\mu \mathcal{P}_{ij}
\mathcal{B}_j) \nonumber \\&& + \frac{1}{2} g_{\mathcal{B^\ast
B^\ast P}}\epsilon_{\mu\nu\alpha\beta} \bar{\mathcal{B}}^{*\mu}_i
\partial^\nu \mathcal{P}_{ij} {\stackrel{\leftrightarrow}{\partial}^\alpha}
\mathcal{B}^{*\beta}_j,
\end{eqnarray}
where $\mathcal{V}$ and $\mathcal{P}$ are $3\times 3$ matrices corresponding to the pseudoscalar and vector octets, which satisfy
\begin{eqnarray}
\mathcal{V}= \left(
  \begin{array}{ccc}
    \frac{\rho^0}{\sqrt{2}}+\frac{\omega_8}{\sqrt{6}} & \rho^+ & K^{*+} \\
    \rho^- &  -\frac{\rho^0}{\sqrt{2}}+\frac{\omega_8}{\sqrt{6}}  & K^{*0} \\
    K^{*-} & \bar{K}^{*0} & -\sqrt{\frac{2}{3}}\omega_8 \\
  \end{array}
\right),
\end{eqnarray}
\begin{eqnarray}
\mathcal{P}= \left(
  \begin{array}{ccc}
    \frac{\pi^0}{\sqrt{2}}+\frac{\eta}{\sqrt{6}} & \pi^+ & K^{+} \\
    \pi^- & -\frac{\pi^0}{\sqrt{2}}+\frac{\eta}{\sqrt{6}}  & K^{0} \\
    K^{-} & \bar{K}^{0} & -\sqrt{\frac{2}{3}}\eta \\
  \end{array}
\right),
\end{eqnarray}
with $\omega_8=\omega\cos\theta+\phi\sin\theta$ and $\sin\theta= -0.761$.
These effective Lagrangians are constructed by considering heavy quark limit and
chiral symmetry. The coupling constants
in the above Lagrangians can be defined as $g_{B^\ast B^\ast \pi}
=g_{B^\ast B \pi}/\sqrt{m_B m_{B^\ast}} =2g/f_\pi$ and $g_{BB
\omega} =g_{B^\ast B^\ast \omega} =\beta g_V/\sqrt{2}$, $f_{B^\ast
B^\ast \rho}/m_{B^\ast} =f_{BB^\ast \rho} =\lambda m_\rho/(\sqrt{2}
f_\pi)$, where $g_V=m_\rho/f_\pi,~\beta=0.9,~
\lambda=0.56~\mathrm{GeV}^{-1}$ and $f_\pi=132~\mathrm{MeV}$.

Using the above Lagrangians, we obtain the decay amplitudes for
$\Upsilon(5S) \to B^{\ast 0} B^- \pi^+$
corresponding to five diagrams shown in Fig. \ref{Fig:ADia} as
\begin{eqnarray*}
\mathcal{M}_a &=& (i)^3 \int \frac{d^4q}{(2 \pi)^4} [g_{\Upsilon
B^\ast B \pi}\epsilon_{\Upsilon}^\mu] [-ig_{B^\ast B \pi} (iq_\rho)]
[-ig_{B^\ast B \pi} (-iq^\nu) \epsilon_{B^\ast}^\nu] \nonumber\\
&&\times \frac{-g^{\mu \rho} + p_1^\mu p_1^\rho/m_{B^\ast}^2}{p_1^2
-m_{B^\ast}^2} \frac{1}{p_2^2 -m_B^2} \frac{1}{q^2 -m_B^2}
\mathcal{F}^2(q^2),\nonumber\\
\mathcal{M}_b &=& (i)^3 \int \frac{d^4q}{(2 \pi)^4} [g_{\Upsilon
B^\ast B \pi}\epsilon_{\Upsilon}^\mu] [-i g_{BB \rho}
(-ip_{1\rho}+ip_{4 \rho})] \nonumber\\ &&\times [ig_{B^\ast B^\ast
\rho} (ip_{5 \beta}-ip_{2\beta}) \epsilon_{B^\ast}^\nu g_{\alpha
\nu} + 4if_{B^\ast B^\ast \rho} (-iq_{\alpha} g_{\nu \beta}
\nonumber\\&& +iq_{\beta} g_{\nu \alpha}) \epsilon_{B^\ast}^\nu]
\frac{1}{p_1^2 -m_B^2} \frac{-g^{\mu \alpha} +p_2^\mu
p_2^\alpha/m_{B^\ast}^2}{p_2^2 -m_{B^\ast}^2} \nonumber\\&& \times
\frac{-g^{\rho \beta} +q^\rho q^\beta/m_\rho^2}{q^2 -m_\rho^2}
\mathcal{F}^2(q^2), \nonumber\\
\mathcal{M}_c &=& (i)^3 \int \frac{d^4q}{(2 \pi)^4} [g_{\Upsilon
B^\ast B \pi}\epsilon_{\Upsilon}^\mu] [-2 f_{B^\ast B \rho}
\varepsilon_{\rho \lambda \alpha \beta} (iq^\rho)
(-ip_1^\alpha-ip_4^\alpha)] \nonumber\\&&\times [-2 f_{B^\ast B
\rho} \varepsilon_{\theta\phi \delta \nu} (-iq^\theta) (ip_5^\delta
+ip_2^\delta) \epsilon_{B^\ast}^\nu] \frac{-g^{\mu \beta} + p_1^\mu
p_1^\beta/m_{B^\ast}^2}{p_1^2 -m_{B^\ast}^2} \nonumber\\&&\times
\frac{1}{p_2^2 -m_B^2 } \frac{-g^{\lambda \phi} + q^\lambda
q^\phi/m_{B^\ast}^2}{q^2 -m_{B^\ast}^2} \mathcal{F}^2(q^2), \nonumber\\
\mathcal{M}_d &=& (i)^3 \int \frac{d^4q}{(2 \pi)^4} [-ig_{\Upsilon
B^\ast B^\ast \pi} \varepsilon_{\mu \nu \alpha \beta}
\epsilon_{\Upsilon}^\mu (ip_3^\alpha) -ih_{\Upsilon B^\ast B^\ast
\pi} \varepsilon_{\alpha \mu \nu \alpha} \nonumber\\
&& \times (-ip_0^\alpha) \epsilon_{\Upsilon}^\mu][-2 f_{B^{\ast- }
B^- \rho^0} \varepsilon_{\delta \tau
\theta \phi} (iq^\delta) (-ip_1^\theta -ip_4^\theta)]\nonumber\\
&&\times [ig_{B^{\ast 0} B^{\ast 0 \rho^0}}((ip_{5 \rho} + ip_{2
\rho}) g_{\lambda \omega} \epsilon_{B^\ast}^\omega) +4i f_{B^{\ast
0} B^{\ast 0} \rho^0} \nonumber\\ && \times (-iq_{\lambda} g_{\rho
\omega} + iq_{\omega} g_{\lambda \rho}) \epsilon_{B^\ast}^\omega]
\frac{-g^{\beta \phi}+ p_1^\beta p_1^\phi/m_{B^\ast}^2}{ p_1^2
-m_{B^\ast}^2} \nonumber\\ && \times \frac{-g^{\nu \lambda}+ p_2^\nu
p_2^\lambda/m_{B^\ast}^2}{p_2^2-m_{B^\ast}^2} \frac{-g^{\tau \rho}+
q^\tau q^\rho/m_{\rho}^2}{q^2 -m_{\rho}^2} \mathcal{F}^2 (q^2),
\nonumber\\
\mathcal{M}_e &=& (i)^3 \int \frac{d^4q}{(2 \pi)^4} [-ig_{\Upsilon
B^\ast B^\ast \pi} \varepsilon_{\mu \nu \alpha \beta}
\epsilon_{\Upsilon}^\mu (ip_3^\alpha) \nonumber\\ && -ih_{\Upsilon
B^\ast B^\ast \pi} \varepsilon_{\alpha \mu \nu \alpha}
(-ip_0^\alpha) \epsilon_{\Upsilon}^\mu][ig_{B^{\ast-} B^- \pi^0}
(-iq^\rho)]\nonumber\\&& \times [-g_{B^{\ast 0} B^{\ast 0} \pi^0}
\varepsilon_{\delta \omega \theta \phi} (ip_5^\delta)
\epsilon_{B^\ast}^\omega (-ip_2^\theta)] \frac{-g^{\beta
\rho}+p_1^\beta p_1^\rho/m_{B^\ast}^2}{
p_1^2-m_{B^\ast}^2} \nonumber\\
&&\times \frac{-g^{\nu \phi} +p_2^\nu
p_2^\phi/m_{B^\ast}^2}{p_2^2-m_{B^\ast}^2} \frac{1}{q^2-m_{\pi}^2}
\mathcal{F}^2 (q^2).
\end{eqnarray*}
Similarly, one also gets the amplitudes corresponding to two
diagrams listed in Fig. \ref{Fig:BDia} as
\begin{eqnarray}
\mathcal{M}_{1} &=&  (i)^3 \int \frac{d^4q}{(2 \pi)^4}
[-ig_{\Upsilon B^\ast B^\ast \pi} \varepsilon_{\mu \lambda \alpha
\beta} \epsilon_{\Upsilon}^\mu (ip_3^\alpha-ip_0^\alpha)]
\nonumber\\
&&\times [-g_{B^\ast B^\ast \pi} \varepsilon_{\delta \tau \theta
\nu} (-ip_1^\delta) (ip_4^\theta) \epsilon_{B^\ast}^\nu] [-g_{B^\ast
B^\ast \pi} \varepsilon_{\phi \rho \zeta \kappa} (ip_5^\phi)
\epsilon_{\bar{B}^\ast}^\rho \nonumber\\
&&\times (-ip_2^\zeta)] \frac{-g^{\lambda \tau} +p_1^\lambda
p_1^\tau/m_{B^\ast}^2}{p_1^2 -m_{B^\ast}^2} \frac{-g^{\beta \kappa}
+p_2^\beta p_2^\kappa/m_{B^\ast}^2 }{p_2^2 -m_{B^\ast}^2} \nonumber\\
&&\times \frac{1}{q^2 -m_{\pi}^2} \mathcal{F}^2(q^2),\nonumber\\
\mathcal{M}_{2} &=&  (i)^3 \int \frac{d^4q}{(2 \pi)^4}
[-ig_{\Upsilon B^\ast B^\ast \pi} \varepsilon_{\mu \lambda \alpha
\beta} \epsilon_{\Upsilon}^\mu (ip_3^\alpha-ip_0^\alpha)] \nonumber\\
&&\times [ig_{B^\ast B^\ast \rho} (ip_{4 \delta}+ip_{1 \delta})
\epsilon_{B^\ast}^\nu g_{\nu \theta}  +4i f_{B^\ast B^\ast \rho}
(iq_{\theta} g_{\nu \delta} \nonumber\\
&& -iq_{\nu}g_{\theta \delta} ) \epsilon_{B^\ast}^\nu] [ig_{B^\ast
B^\ast \rho} (-ip_{2\tau} -ip_{5 \tau}) \epsilon_{\bar{B}^\ast}^\rho
g_{\rho \zeta} \nonumber\\&& +4if_{B^\ast B^\ast \rho} (-iq_\zeta
g_{\rho \tau} + iq_\rho g_{\zeta \tau}) \epsilon_{\bar{B}^\ast}^\rho
] \frac{-g^{\lambda \theta} +p_1^\lambda
p_1^\theta/m_{B^\ast}^2}{p_1^2-m_{B^\ast}^2} \nonumber\\
&&\times   \frac{-g^{\beta \zeta}+ p_2^\beta
p_2^\zeta/m_{B^\ast}^2}{ p_2^2 -m_{B^\ast}^2 } \frac{-g^{\delta
\tau} +q^\delta q^\tau/m_{\rho}^2 }{q^2 -m_{\rho}^2 }
\mathcal{F}^2(q^2).\nonumber
\end{eqnarray}
In these expressions for decay amplitudes, the dipole form factor
(FF)
$$\mathcal{F}(q^2)=\left(\frac{\Lambda^2-m^2}{q^2-m^2}\right)^2$$ is
introduced to describe the structure effect of the
$B^{(*)}B^{(*)}\pi$ and $B^{(*)}B^{(*)}\rho$ interaction vertexes in
Figs. \ref{Fig:ADia} and \ref{Fig:BDia}. The parameter $\Lambda$
introduced in the FF can be parameterized as
$\Lambda=m+\alpha\Lambda_{QCD}$ with $\Lambda_{QCD}=220$ MeV with a
new parameter $\alpha$, where $m$ denotes the mass of the exchanged
light meson.

The total decay amplitudes are expressed as
\begin{eqnarray}
\mathcal{A}_{1}&=&\mathcal{M}_a + \mathcal{M}_b +\mathcal{M}_c,\\
\mathcal{A}_{2}&=&\mathcal{M}_d +\mathcal{M}_e,\\
\mathcal{A}_{3}&=&\mathcal{M}_1 +\mathcal{M}_2,
\end{eqnarray}
where $\mathcal{A}_1$ and $\mathcal{A}_2$ correspond to $\Upsilon(5S)\to B^{*0} B^- \pi^+$ with the intermediate $B\bar{B}^*+h.c.$ and $B^*\bar{B}^*$, respectively, while $\mathcal{A}_{3}$ to $\Upsilon(5S)\to B^{*0} B^{*-} \pi^+$ with the intermediate $B^*\bar{B}^*$.
The general differential decay width for
$\Upsilon(5S)(p_0) \to \pi(p_3) B^{(\ast)}(p_4) B^\ast (p_5) $ is
\begin{eqnarray}
d\Gamma_i =\frac{1}{3} \frac{1}{(2 \pi)^3} \frac{1}{32
m^3_{\Upsilon(5S)}} \overline{|\mathcal{A}_i|^2} dm_{B^{\ast}
B^{(\ast)}}^2 dm_{B^\ast \pi}^2 \label{t1}\quad (i=1,2,3)
\end{eqnarray}
with $m_{B^{\ast} B^{(\ast)}}^2 =(p_4 +p_5)^2$ and $m_{B^\ast
\pi}^2= (p_3 +p_5)^2$, where the overline indicates the sum over the
polarization of $\Upsilon(5S)$ in the initial state and the
polarizations of $B^\ast$ or $\bar{B}^\ast$ meson in the final
state.

Since we mainly concentrate on the lineshapes of the $B\bar{B}^*$
and $B^*\bar{B}^*$ invariant mass spectrum distributions of
$\Upsilon(5S)\to B^{*0} B^- \pi^+$ and $\Upsilon(5S)\to B^{*0}
B^{*-} \pi^+$ decays, the interference effects between
$\mathcal{A}_1$ and $\mathcal{A}_2$ are not considered in this work.
Calculating the distributions of Eq. (\ref{t1}), we can see whether
there exist the enhancement structures close to the $B\bar{B}^*$ and
$B^*\bar{B}^*$ thresholds steming from the ISPE mechanism. As one
can see from Fig. (3), peaks of our theoretical curves nicely match
with those of the experimental enhancement structures.

\begin{figure}[htbp]
\centering%
\scalebox{0.6}{\includegraphics{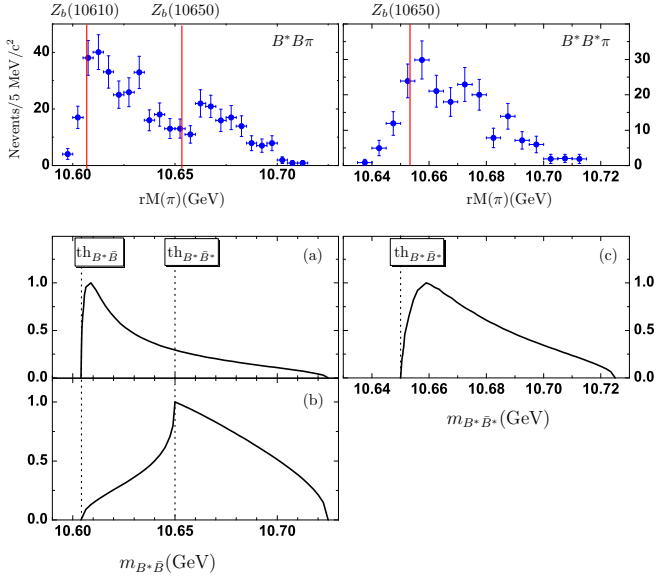}}%
\caption{(Color Online) The theoretical curves for distributions of
$d\Gamma(\Upsilon(5S) \to B^{\ast 0} B^- \pi^+)/dm_{B\bar{B}^*}$
(left) and $d\Gamma(\Upsilon(5S) \to B^{\ast 0} B^{*-}
\pi^+)/dm_{B^* \bar{B}^*}$ (right) (see diagrams (a)-(c)). Here, the
maximum of the theoretical line shape is normalized to be 1 and the
typical value of $\alpha=1$ is taken in our calculation. Diagrams
(a) and (b) correspond to $\Upsilon(5S) \to B^{\ast 0} B^- \pi^+$
via the intermediates $B\bar{B}^*+h.c.$ and $B^*\bar{B}^*$,
respectively, by the ISPE mechanism. The diagram (c) reflects the
distribution $d\Gamma(\Upsilon(5S) \to B^{\ast 0} B^{*-}
\pi^+)/dm_{B^* \bar{B}^*}$ of $\Upsilon(5S) \to B^{\ast 0} B^{*-}
\pi^+$. To compare our theoretical results with the experimental
data, we also show Belle's data (the blue dots with error) of the
$\Upsilon(5S)\to B\bar{B}^*\pi$ (left) and $\Upsilon(5S)\to
B^*\bar{B}^*\pi$ (right) \cite{newbelle}. The thresholds of
$B\bar{B}^*$ and $B^*\bar{B}^*$ are marked by the dotted lines.
\label{Fig:openbottom}}
\end{figure}

\section{The open-charm decays of higher charmonia with a single pion emission}\label{sec3}

Being inspired by the success of the former section, we would like
to apply the ISPE mechanism to the open charm decays of higher
charmonia with a single pion emission, for example, to the processes
$\psi(4415)\to \pi^+D^{*0} D^{(\ast)-}$ and $\psi(4160)\to
\pi^+D^{*0} D^{-}$.

What we need in this section is to replace $\Upsilon(5S)$, $B$, and
$B^*$ in Figs. 1 and 2 with $\psi(4415)/\psi(4160)$, $D$, and $D^*$,
respectively. We also need to replace the corresponding fields in
the effective Lagrangians in Eqs. (1-3). The parameters are of
course new definitions. The resultant curves are shown in Fig. (4).
Similarly to $\Upsilon(5S) \to \pi B^{\ast}
\bar{B}^{(\ast)}$, there are two significant enhancement structures
near the thresholds of $D^\ast \bar{D}$ and $D^\ast \bar{D}^\ast$ in
the invariant mass
spectra $m_{D^*\bar D}$ and $m_{D^*\bar D^*}$ of $\psi(4415) \to \pi D^\ast \bar{D}^{(\ast)}$. For
$\psi(4160) \to \pi D^\ast \bar{D}$, only one enhancement
near $D^\ast \bar{D}$ threshold is predicted and the threshold of
$D^\ast \bar{D}^\ast$ is out of the range of the invariant mass
spectra $m_{D^*\bar D}$ of this process.

\begin{figure}[htbp]
\centering%
\scalebox{0.7}{\includegraphics{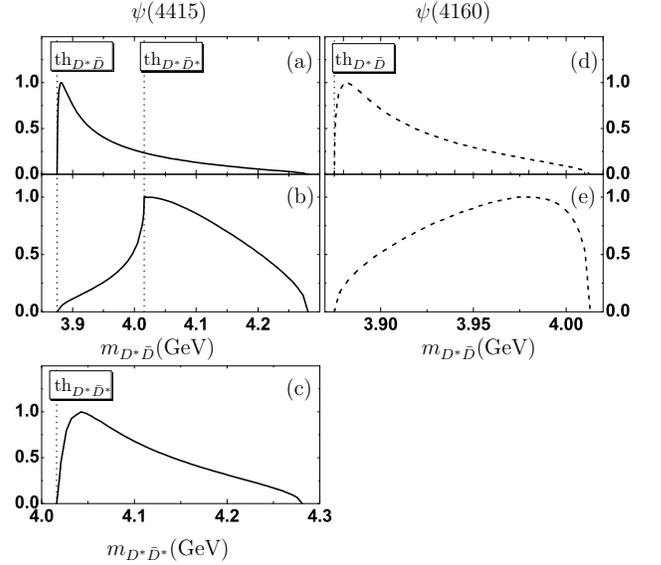}}%
\caption{The theoretical curves for $D^\ast \bar{D}^{(\ast)}$
invariant mass spectrum in open charm pion decays of higher charmonia $\psi(4415)$ and
$\psi(4160)$. The curves for
$d\Gamma(\psi(4415) \to D^{\ast 0} D^{-} \pi^+)/d{m_{D^\ast
\bar{D}}}$ via $D^{\ast 0} D^-$ and $D^{\ast 0} D^{\ast -}$ in the
ISPE frame correspond to diagrams (a) and (b), respectively.
The diagram (c) indicates the ISPE mechanism predictions for
$d\Gamma(\psi(4415) \to D^{\ast 0} D^{\ast -} \pi^+)/d{m_{D^\ast
\bar{D}^\ast}}$ via $D^{\ast 0} D^{\ast -}$ intermediate. For
$\psi(4160)$ only $D^\ast D \pi$ process is allowed, and diagrams
(d) and (e) express $d\Gamma(\psi(4160) \to D^{\ast 0} D^{-}
\pi^+)/d{m_{D^\ast \bar{D}}}$ via $D^{\ast 0} D^-$ and $D^{\ast 0}
D^{\ast -}$, respectively.  \label{Fig:charm}}
\end{figure}

\section{Summary}\label{sec4}

Very recently, the Belle Collaboration has reported new results on
$Z_b(10610)$ and $Z_b(10650)$ at the ICHEP2012 conference that these
$Z_b$ structures also exist in the $B\bar{B}^*$ and $B^*\bar{B}^*$
invariant mass spectra of $\Upsilon(5S)\to \pi B\bar{B}^*,\,\pi
B^*\bar{B}^*$ decays \cite{newbelle}. This motivates us to apply the
ISPE mechanism because these are the typical processes for this
mechanism to be applied. Using the effective Lagrangian approach
among hadrons as well as chiral particles, we have computed the
theoretical curves of the invariant mass spectra of $B^{\ast} \bar
B^{(*)}$ for the above processes, which are shown in Figs. 3 and
have successful agreement with experimental enhancement structures
of $Z_b(10610)$ and $Z_b(10650)$.

This success has further driven us to apply the ISPE mechanism to
the open-charm decays of higher charmonia with a single pion
emission, $\psi(4415)\to \pi^+D^{*0} D^-$ and $\psi(4160)\to
\pi^+D^{*0} D^{-}$. Similar procedures to those in Sec. II have led
us to depict the theoretical curves of the invariant mass spectra as
shown in Fig. 4. Figure 4 shows two clear peaks for the invariant mass
spectra $m_{D^*\bar D}$ and $m_{D^*\bar D^*}$ of the decay
$\psi(4415)\to \pi^+D^{*0} D^-$ and one peak for $m_{D^*\bar D}$ of
$\psi(4160)\to \pi^+D^{*0} D^{-}$. These predictions can be easily
tested by Bell, BaBar, forthcoming BellII and SuperB.

\section*{Acknowledgment}

This project is supported by the National Natural Science
Foundation of China under Grants Nos. 11175073, 11005129, 11035006, the Ministry of Education of China (FANEDD
under Grant No. 200924, DPFIHE under Grant No. 20090211120029,
NCET under Grant No. NCET-10-0442, the Fundamental Research Funds
for the Central Universities), and the West Doctoral Project of
Chinese Academy of Sciences.

\end{document}